\documentclass[english,3p,number,sort&compress,times]{elsarticle}
\usepackage[T1]{fontenc}
\usepackage[latin9]{inputenc}
\usepackage{amsmath}
\usepackage{amssymb}
\usepackage{ifthen}
\usepackage{graphicx}
\usepackage{times}
\usepackage{babel}
\usepackage{color}
\usepackage{bm}

\newcommand{\ifis}[2]{
\ifthenelse{\equal{#1}{}}{}{#2}
}

\def\be{\begin{equation}}
\def\ee{\end{equation}}
\def\bea{\begin{eqnarray}}
\def\eea{\end{eqnarray}}
\def\bsu{\begin{subequations}}
\def\esu{\end{subequations}}
\def\bi{\begin{itemize}}
\def\ei{\end{itemize}}

\newcommand{\op}[1]{\widehat{#1}}
\newcommand{\dagop}[1]{\widehat{#1}^{\dagger}}
\newcommand{\bo}[1]{{\mathbf{#1}}}
\newcommand{\mc}[1]{{\mathcal{#1}}}
\newcommand{\wt}[1]{{\widetilde{#1}}}
\newcommand{\wb}[1]{{\overline{#1}}}
\newcommand{\ul}[1]{{\underline{#1}}}

\newlength{\templength}

\newenvironment{eq}{\begin{equation}}{\end{equation}}

\newcommand{\REM}[1]{\ifthenelse{0=1}{#1}{}}

\begin{document}


\title{A tractable prescription for large-scale free flight expansion of wavefunctions}

\author{P. ~Deuar} 
\ead{deuar@ifpan.edu.pl}
\address{Institute of Physics, Polish Academy of Sciences, Al. Lotnik\'ow 32/46, 02-668 Warsaw, Poland}

\begin{abstract}
A numerical recipe is given for obtaining the density image of an initially compact quantum mechanical wavefunction that has expanded 
by a large but finite factor under free flight. 
The recipe given avoids the memory storage problems that plague this type of calculation by reducing the problem to the sum of a number of fast Fourier transforms carried out on the relatively small initial lattice. The final expanded state is given exactly on a coarser magnified grid with the same number of points as the initial state. 
An important application of this technique is the simulation of measured time-of-flight images in ultracold atom experiments, especially when the initial clouds contain superfluid defects. 
It is shown that such a finite-time expansion, rather than a far-field approximation is essential to correctly predict images of defect-laden clouds, even for long flight times. Examples shown are: an expanding quasicondensate with soliton defects and a matter-wave interferometer in 3D.
\end{abstract}


\maketitle

\section{Introduction}

The free flight expansion of a quantum wavefunction, though physically very simple, is often a troublesome computational problem if the state that is required is not quite yet in the far field regime. 
The snag is that a computational lattice that both resolves the small initial cloud and encompasses the large expanded cloud can be prohibitively large. 
Here, it will be described how to overcome this while still using standard discrete fast Fourier transform (FFT) tools. 

For example, this is commonly desired when simulating experiments in ultracold atoms. A ubiquitous experimental procedure in this field is the release of the atoms from a trap and the subsequent observation of the density of a strongly expanded cloud. Given that the imaged expanded cloud is usually much larger than the initial one pre-release, the observed expanded atom density corresponds approximately to the velocity distribution in the initial cloud. More precisely, it corresponds to the velocity distribution that is formed early on after release, when the interatomic interaction energy has been converted into kinetic energy. This is the picture that is often used to interpret the data. 

This interpretation assumes that the detection is occurring in the far field where all structure is large compared to the initial cloud. However, in practice this is often not a good enough approximation, particularly if one is interested in fine structure inside the atomic cloud, such as defects or interparticle correlations. The reality is that the expansion is usually by a factor of tens or hundreds, so that interesting features such as defects or correlations that are of the order of 10\% or 1\% of the initial cloud in size have not yet yet attained a velocity profile at the time of detection. They are already distorted from their spatial profile in the initial cloud, but their shape has not yet stabilized to its far field form. 
Some examples where a long but not quite far-field expansion occurs include the interference pattern generated after release of a pair of elongated clouds \cite{Gring12,AduSmith13}, the study of Hanbury Brown-Twiss correlations in expanding clouds \cite{Perrin12,Gawryluk15} and two-particle correlations in a halo of supersonically scattered atoms \cite{Jaskula10,Kheruntsyan12}. 

The basic numerical task here is to predict the detected density image based on whatever model we are using for the atomic field. For excited or thermal gases an ensemble of classical field \cite{Kagan97,Davis01,Goral01,Brewczyk07,Proukakis08,Blakie08,book13} or truncated Wigner wavefunctions \cite{Steel98,Sinatra02,Polkovnikov10,Martin10a} are often used. 
The straightforward approach is to place the whole field $\Psi(\bo{x},t)$ from the outset in an x-space large enough to accommodate the whole expansion. 
However, it is rarely technically feasible to carry out the entire expansion by this method in three dimensions despite the seemingly trivial physics. A good description of the initial state in a three-dimensional lattice can easily require $\mc{O}(10^5)$ lattice points or more, and an expansion by a factor of 10--100 in each direction would lead to $10^8 - 10^{11}$ lattice points. This is either intractable or impractical on a simple computer, and even more so for studies of defect statistics or correlations which require surveys of hundreds or thousands of realizations.

Barring access to supercomputing resources, a standard resort in this case is to make a somewhat unsavory compromise: Simulate the expansion as far in time as the computer allows, and assume that the neglected later changes are not qualitatively important. It will be shown here how to avoid this while still using standard discrete FFT tools on a simple computer. 

In Sec.~\ref{MATTER} the basics of the problem are described, and in Sec.~\ref{DEFECTS} an estimate is made of the the time regime over which a careful exact expansion of clouds with defects is necessary. Sec.~\ref{1DEX} demonstrates this with an example. The numerical difficulty is examined in more detail in Sec.~\ref{NUM}, and the solution derived in Sec.~\ref{DERIV}. The paper concludes in Sec~\ref{DISCUSSION} with some discussion of practicalities and applications. 

The prescription that constitutes the main result is briefly given in a stand-alone form in Sec.~\ref{RESULT}.

\section{The matter at hand}
\label{MATTER}

The aim is to calculate what is actually measured, the spatial density distribution at the detector, $\rho(\bo{r},t_{\rm final})$. We assume that just before the trap is switched off at $t=0$, the trapped state is described by a complex field $\Psi_0(\bo{r})$ that has the form of a single-particle wavefunction.

\subsection{Conversion phase}
\label{CONVERSION}

Typically the expansion can be considered as consisting of two phases: An initial ``conversion'' phase during which the interaction energy between the atoms is converted to kinetic energy, and later free flight of the atoms. 
Since the interaction energy per particle is proportional to the density, an expansion in three-dimensional space by a factor of two in size will reduce this interaction energy per particle by a factor of eight. 
This initial expansion can be done in a straightforward way until interactions are diluted away to become negligible. One just takes a computational lattice $\wb{x}$ in x-space that is 2--4 times wider than the initial cloud $\Psi_0$, and evolves on that. In ultracold atoms, the workhorse Gross-Pitaevskii Equation (GPE) is typically used -- see the example in Sec.~\ref{3DEX}. 

The end result of this phase (at time $t_s$, say) is that we have a partly expanded wavefunction $\Psi(\bo{r},t_s)$. Numerically, it is described as a table of complex numbers $\Psi_{\bo{n}}$ indexed by the set of integers $\bo{n} =\{n_1,\dots,n_d\}$ in $d$-dimensional space, that enumerate the points on the numerical lattice.  
The lattice spacings are $\Delta x_j = L_j/M_j$ where $L_j$ is the length of the box in the $j$th direction, and $M_j$ the corresponding number of lattice points, so that $n_j = 0,1,\dots,(M_j-1)$. The lattice positions are 
\be\label{xj}
x_j = a_j + \Delta x_j n_j
\ee
with offsets $a_j$. i.e. $\bo{x} = \bo{a}+\Delta\bo{x}\cdot\bo{n}$.

Note that the wavefunctions $\Psi(\bo{r})$ are not in general the complete quantum many-particle wavefunction unless the particles are non-interacting. For interacting particles, one usually works with $\Psi$ in some kind of c-field approximation \cite{Brewczyk07,Proukakis08,Blakie08,book13,Sinatra02,Polkovnikov10}.

\subsection{Lattice notation}
Several numerical lattices will appear in what follows. The following notation will be used: 
\begin{itemize}
\item Quantities with a tilde, such as $\wt{\Psi}(\bo{k})$, are in k-space.
\item Bold quantities are vectors in $d$ dimensions (usually $d=3$), with elements indexed by $j$ as in $\bo{x}=\{x_1,\dots,x_j\dots, x_d\}$.
\item Undecorated quantities, such as $\Psi(\bo{x})$ denote the lattice used to represent the starting state at $t_s$. This has a manageable number of points, $M$.
\item Barred quantities, such as $\wb{\Psi}(\wb{\bo{x}})$ will be on a magnified numerical lattice $\wb{\bo{x}}$ that can describe the expanded state, but is too coarse to describe the starting state at $t_s$.
\item Underlined quantities such as $\ul{\Psi}(\ul{\bo{x}})$ will be used to denote a sufficiently huge lattice that both encompasses the expanded state at $t_{\rm final}$ and resolves the starting state at $t_s$, when this lattice is different from the starting undecorated one. 
\item The position coordinate $\bo{r}$ is a continuum quantity, as opposed to $\bo{x}$ which are corresponding lattice positions. Similarly, $\bm{\kappa}$ is a continuum wavevector, while $\bo{k}$ is discretized.
\end{itemize}

\subsection{Free flight into the far field}
\label{FREEFLIGHT}

The remaining evolution after $t_s$, the ``starting time'', is just free flight. Each particle of momentum $\hbar\bm{\kappa}$ flies a distance $\hbar\bm{\kappa}\,t_{\rm flight}/m$, 
where the flight time is 
\be
t_{\rm flight} = t_{\rm final} - t_s\ .
\ee
Then with a far field assumption, i.e. that the initial starting position at $t_s$ is irrelevant because they have flown so far, the position of a particle is 
\be
\bo{r} = \hbar t_{\rm flight} \bm{\kappa}/m.
\ee
An  estimate for the final density can then be obtained from $\wt{\Psi}(\bm{\kappa},t_s)$, the momentum-space wavefunction at the end of the conversion phase. It is:
\be\label{farfieldk}
\rho_{\rm ff}(\bo{r}) = |\Psi_{\rm ff}(\bo{r})|^2 = \left(\frac{m}{\hbar t_{\rm flight}}\right)^{d}\,\left|\,\wt{\Psi}\left(\frac{m\bo{r}}{\hbar t_{\rm flight}},t_s\right)\,\right|^2.
\ee
The prefactor is for normalization purposes, so that $\int d^d\bm{\kappa}\, |\wt{\Psi}(\bm{\kappa})|^2 = \int d^d\bo{r}\, |\Psi_{\rm ff}(\bo{r})|^2$.
Notably, this discards any phase information. However, the usual imaging in ultracold atom experiments is insensitive to that.

The starting momentum wavefunction $\wt{\Psi}(\bm{\kappa})$ at $t_s$ is obtained with a norm preserving Fourier transform:
\be\label{FT}
\wt{\Psi}(\bm{\kappa}) = \frac{1}{(2\pi)^{d/2}}\int d^d\bo{r}\,e^{-i\bm{\kappa}\cdot\bo{r}}\,\Psi(\bo{r}).
\ee
Numerically, the conversion is best made with a discrete Fourier transform (DFT). The DFT of a field $Q$ is 
\be\label{DFT}
\wt{Q}_{\wt{\bo{m}}} = {\rm DFT}\left[ Q_{\bo{n}} \right]_{\wt{\bo{m}}} = \sum_{\bo{n}} Q_{\bo{n}} \exp\left[-i\sum_{j=1}^d\frac{2\pi}{M_j} n_j \wt{m}_j \right].
\ee
with indices $\wt{m}_j=0,1,\dots,(M_j-1)$ labeling the position on the k-space lattice which has spacing $\Delta k_j = 2\pi/L_j$. The sum is over the whole $\bo{n}$ range.

In what follows, we will always be using the physical free-space wavevectors $\bo{k}_{\wt{\bo{m}}}$ ordered as:
\be\label{kj}
k_j(\wt{m}_j) = \Delta k_j \wt{l}_j = \Delta k_j \times \left\{\begin{array}{l@{\quad{\rm for\ \ }}l}
\wt{m}_j&\wt{m}_j<M_j/2\\
\wt{m}_j-M_j&\wt{m}_j\ge M_j/2
\end{array}\right.
\ee
The integer multipliers can also be written as  $\wt{l}_j = {\rm mod}\left[\wt{m}_j+\frac{1}{2}M_j\,,\, M_j\right]-\frac{1}{2}M_j$.
For simple transformations such as (\ref{FT}) and (\ref{Psiwtm}), a set of monotonically ordered non-negative wavevectors $\wt{\bo{m}}\cdot\Delta\bo{k}$ is equivalent operationally to (\ref{kj}) because $\Delta k_j M_j(x_j-a_j)$ is an integer multiple of $2\pi$. However, such equivalence no longer holds for 
calculating the kinetic energy or upon changing the lattice offset $\bo{a}$, both of which will be needed for expansion.

Using (\ref{FT}) and the DFT (\ref{DFT}), as well as taking care of a possible offset in (\ref{xj}), the discrete momentum wavefunction at $t_s$ is
\be\label{Psiwtm}
\wt{\Psi}_{\wt{\bo{m}}}(t_s) = \frac{\Delta V}{(2\pi)^{d/2}} e^{-i\bo{a}\cdot\bo{k}_{\wt{\bo{m}}}}\ {\rm DFT}\left[\Psi_{\bo{n}}(t_s)\right]_{\wt{\bo{m}}}.
\ee
with lattice point volume $\Delta V=\prod_j\Delta x_j$. This is readily implemented using standard fast Fourier transform (FFT) libraries \cite{FFTW05,FFTW98}.

In most cases in the literature, the short initial expansion to $t_s$ and conversion (\ref{farfieldk}) to obtain the detected density is all that is done. 
This is fine provided that we are only interested in momentum differences $\delta\bo{k}$ much larger than those corresponding to the width $W_s$ of the converted cloud at $t_s$. That is, when $|\delta \bo{k}| \gg 
m W_s/\hbar t_{\rm flight}$. Or, alternatively, that we are only interested in spatial resolutions $\gg W_s$ in the final expanded cloud.

\subsection{Free flight without a far field assumption}
\label{KINETIC}
To avoid making the rather uncontrolled far field assumption (\ref{farfieldk}), and get results for a well defined final time $t_{\rm final}$, consider first that in principle, the free flight evolution of the wavefunction in k-space is straightforward:
\be\label{free0}
\wt{\Psi}(\bo{k},t_{\rm final}) = \wt{\Psi}(\bo{k},t_s) \exp\left[-it_{\rm flight}\frac{\hbar|\bo{k}|^2}{2m}\right].  
\ee
In principle, all one then needs to obtain $\rho(\bo{r},t_{\rm final})$ is to inverse Fourier transform $\wt{\Psi}(\bo{k},t_{\rm final})$ back into x-space. Generally:
\be\label{iFT}
\Psi(\bo{r},t_{\rm final}) = \frac{1}{(2\pi)^{d/2}}\int d^d\bm{\kappa}\,e^{i\bm{\kappa}\cdot\bo{r}}\ \wt{\Psi}(\bm{\kappa},t_{\rm final}).
\ee
The discrete implementation like in  (\ref{DFT}) and (\ref{Psiwtm}) is 
\be\label{iDFT}
Q_{\bo{n}} = {\rm DFT}^{-1}\left[ Q_{\wt{\bo{m}}} \right]_{\bo{n}} = \frac{1}{M}\sum_{\wt{\bo{m}}} \wt{Q}_{\wt{\bo{m}}} \exp\left[i\sum_{j=1}^d\frac{2\pi}{M_j} n_j \wt{m}_j \right].
\ee
$M=\prod_jM_j$ is the overall lattice size. With volume $V=\prod_jL_j$, 
\be\label{Psin}
\Psi_{\bo{n}}(t_{\rm final}) = \frac{(2\pi)^{d/2}}{V} {\rm DFT}^{-1}\left[\wt{\Psi}_{\wt{\bo{m}}}(t_{\rm final})e^{i\bo{a}\cdot\bo{k}_{\wt{\bo{m}}}} \right]_{\bo{n}}.
\ee
This step can, however, be hard on computational resources, even with an FFT because a very large lattice $M$ is often needed to fully describe the final time state $\wt{\Psi}(t_{\rm final})$.

\subsection{Continued defect dynamics during free-field expansion}
\label{DEFECTS}
Expansions of clouds containing narrow mobile defects are a popular experimental topic in recent years \cite{Gring12,Chomaz15,Serafini15,Lamporesi13,Donadello14,Sadler06,Weiler08}.
These are systems for which the transition from the starting $t_s$ state to the far-field is nontrivial. 

Let the overall width of the cloud at $t_s$ in a chosen direction be $W_s$, and consider defects of width ${\rm w}_{\rm def}\ll W_s$ and typical speed $u$. Speed differences between different defects are then also $\approx u$. The velocity distribution in the gas, however, is dominated by the shortest length scale in the system. This is usually given by the half-width of individual defects, giving a typical velocity $\sigma_v\approx 2\hbar/m{\rm w}_{\rm def}$. After significant expansion, the width of the cloud will be $W_{\rm final}\approx 2t_{\rm flight}\sigma_v = 4\hbar t_{\rm flight}/m{\rm w}_{\rm def}$.  

It takes a time $t_v=W_s/\sigma_v$ for a rough semblance of the velocity/momentum distribution to form in real space due to expansion  (this is the time for a typical particle to move across the initial cloud). 
Remnants of defects can continue to rearrange until a time when their relative speed would allow them to move as far as $\approx W_s$, which is a typical initial spacing between them:
\be\label{tarrange}
t_{\rm arrange}\approx \frac{W_s}{u} \gg t_{\rm v}.
\ee
For clearly recognizable defects to be present, one should have defects slower than particles: $u\ll \sigma_v$. Due to this slowness, there is a period $t_v\ll t_{\rm flight} \ll t_{\rm arrange}$ during which complicated rearrangement of defect remnants can take place even though the gross shape of the cloud already resembles the far-field velocity distribution. The simple far field expression (\ref{farfieldk}) is not appropriate during this time. 

This is not an uncommon situation in ultracold atom experiments, and has relevance for interpretation of experimental data. For an initially trapped thermal gas in a classical field regime  where its dynamics is quite well described by the Gross-Pitaevskii equation \cite{Sinatra02,Mora03,Castin04,Brewczyk07,Karpiuk12,Pietraszewicz15}, typical defects are solitons in 1D and vortices in 2D. In this regime, when $g$ is the s-wave scattering length and $\rho$ the typical density, the chemical potential is $\mu\approx g \rho$, giving defect width ${\rm w}_{\rm def}\approx2\hbar/\sqrt{m\mu}$ and a speed of sound $c=\sqrt{\mu/m}$. Major defects are much slower, i.e. $u=\epsilon c$ with $\epsilon\ll1$. With a trap frequency of $\omega$, the initial cloud width is $W_s\approx (2/\omega)\sqrt{2\mu/m}$. This lets us estimate $t_v=\sqrt{8}/\omega$ and $t_{\rm arrange}$, and one finds that the times $t_{\rm flight}$ during which rearrangement is still taking place in a cloud that looks to be already far-field in its gross features is 
\be
t_v \lesssim t_{\rm flight} \lesssim t_{\rm arrange}, \qquad\text{i.e.}\qquad 
1 < \frac{\omega t_{\rm flight}}{\sqrt{8}}\lesssim \frac{1}{\epsilon}.
\ee
This can be a significant period. 

\section{Example: soliton dynamics during expansion}
\label{1DEX}

Let us consider an example of complicated free evolution even at times that would naively be considered to be in the far-field: the expansion of an elongated 1D ultracold gas in the quasi-1D regime. 
We take physical parameters like in a series of recent experiments \cite{Gring12,AduSmith13}, where clouds in the classical field regime were prepared. 
An initial c-field state of the 1d system is generated at temperature $T=80$nK = $260\hbar\omega/k_B$ using the stochastic Gross-Pitaevskii equation
\be
\hbar\frac{\partial\Psi(x,\tau)}{\partial\tau} = -i(1-i\gamma)\left[-\frac{\hbar^2}{2m}\nabla^2+g_1|\Psi(x,\tau)|^2-\mu\right]\Psi(x,\tau) 
+ \sqrt{2\gamma \hbar k_BT}\ \eta(x,\tau)\label{SGPE}
\ee
by taking a sample of the field $\Psi_{\rm ic}(x) = \Psi(x,\tau)$ at $\tau = 10/\omega$, once the ergodic ensemble is reached. The simulation grows the field from the vacuum $\psi(x,0)=0$. 
Here, $g_1=0.54\hbar\omega a_{\rm ho}$ is the 1D s-wave scattering length for ${}^{87}$Rb in terms of the harmonic oscillator length $a_{\rm ho}=\sqrt{\hbar/m\omega}$. The bath coupling $\gamma=0.02$ has a typical value, $\mu=90\hbar\omega$ is a chemical potential chosen to give $N=3000$ particles on average in the stationary ensemble, and $\eta(x,\tau)$ is a Gaussian complex white noise field with variance $\langle\eta(x,\tau)^*\eta(x',\tau')\rangle = \delta(x-x')\delta(\tau-\tau')$. The lattice cutoff in a plane-wave basis is chosen as $\hbar k_{\rm max}= 0.65\sqrt{2\pi mk_B T}$, according to the optimum values obtained in \cite{Pietraszewicz15}. The generation of $\Psi_{\rm ic}(x)$ was carried out using (\ref{SGPE}) on an initial  lattice with $M=2^{11}$ points and $L=60a_{\rm ho}$.

\begin{figure}[htb]
\begin{center}
\includegraphics[width=\columnwidth]{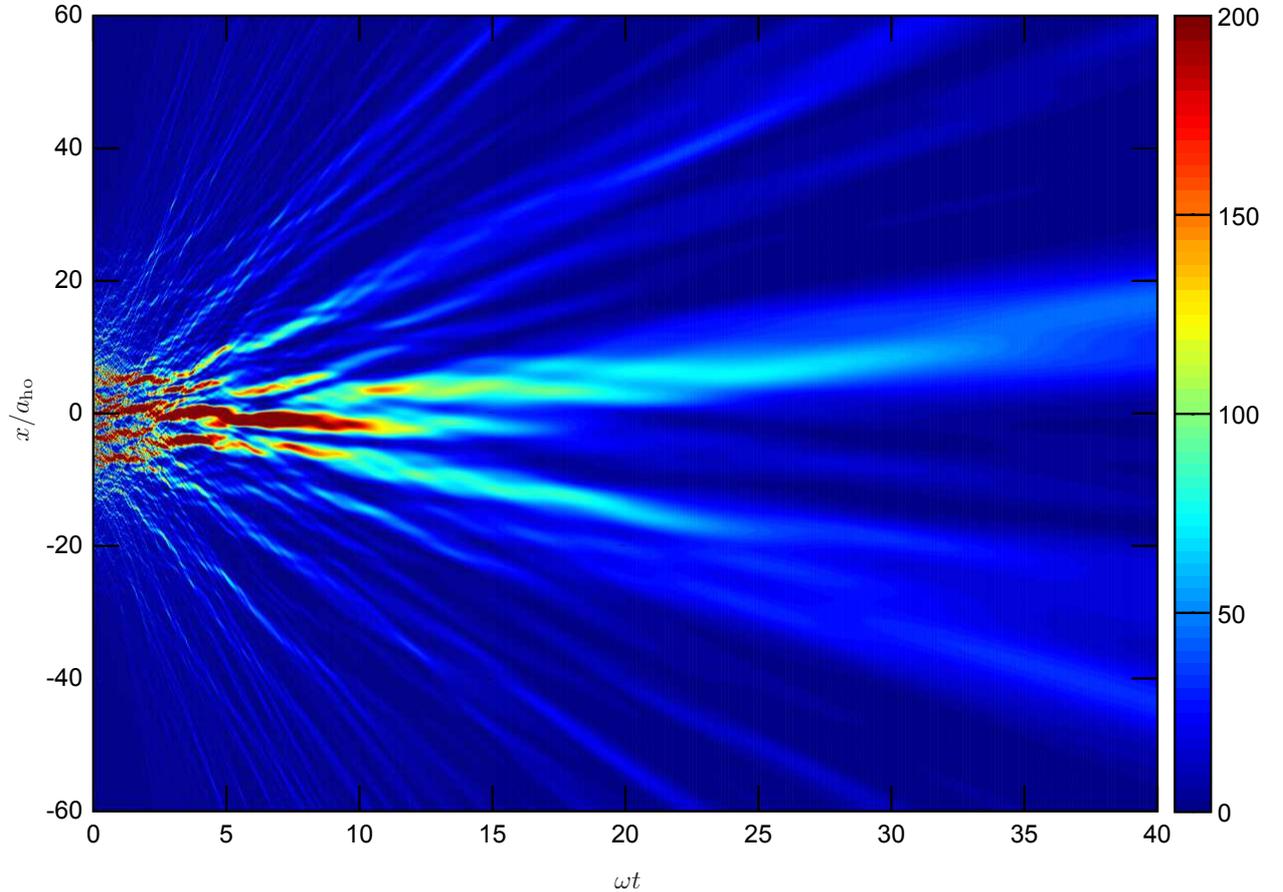}
\end{center}\vspace*{-4mm}
\caption{Evolution of the density $\rho(x)=|\Psi(x)|^2$ (in units of $a_{\rm ho}^{-1}$) after release, calculated directly according to (\ref{freex}), (\ref{Psiwtm}), and (\ref{Psin}). 
The vertical scale is narrowed down compared to the computational lattice, with $\ul{L}=2400a_{\rm ho}$, to show the most interesting region. 
\label{FIG-1}}
\end{figure}

A proper treatment of the conversion phase will be done in the 3D example~\ref{3DEX}. Here, let us just do an immediate free-field expansion of $\Psi_{\rm ic}(x)$ from the moment the trap is switched off at $t_s=t_0=0$.
The 1D density $\rho(x)=|\Psi(x)|^2$ approximates the marginal density of the 3D cloud when integrated over transverse directions. 
The fully free expansion can be quite a good approximation for a very tight initial trap that has the initial gas in a quasi-1d regime (tight transverse trap frequency $\omega_{\perp})$. Release causes a very rapid expansion in the transverse directions on a timescale of $1/\omega_{\perp}$, with width $\propto (1/\omega_{\perp})\sqrt{1+\omega_{\perp}^2t^2}$\, \cite{Castin96}. Accordingly, the density drops as $\sim 1/(1+\omega_{\perp}^2t^2)$, and so does the relative strength of interparticle interactions. After a time of several $1/\omega_{\perp}$ (short compared to $t_v$), the gas is effectively non-interacting.

The evolution of the field is shown in Fig.~\ref{FIG-1}. 
Here in 1D, it is easily done directly using the equation
\be\label{freex}
\frac{\partial\Psi(x,t)}{\partial t} = i\frac{\hbar}{2m}\nabla^2\Psi(x,t) 
\ee
and the DFTs (\ref{Psiwtm},\ref{Psin}). The initial state $\Psi_{\rm ic}(x)$ was padded with vacuum and evolved on a lattice of $\ul{M}=81920$ points on a simulation region of length $\ul{L}=2400a_{\rm ho}$, with $\ul{a}=-\ul{L}/2$.
The purpose is to see defect evolution during expansion. Indeed, we see that appreciable defect evolution occurs until times of about $\approx 30/\omega$. This can be compared to the values of the crude estimates of (\ref{tarrange}) obtained for this system when taking $\epsilon=0.1$: $t_v=2.8$ for significant expansion and $t_{\rm arrange}=28$ for end of rearrangement. A very good match.

\begin{figure}[htb]
\begin{center}
\includegraphics[width=\columnwidth]{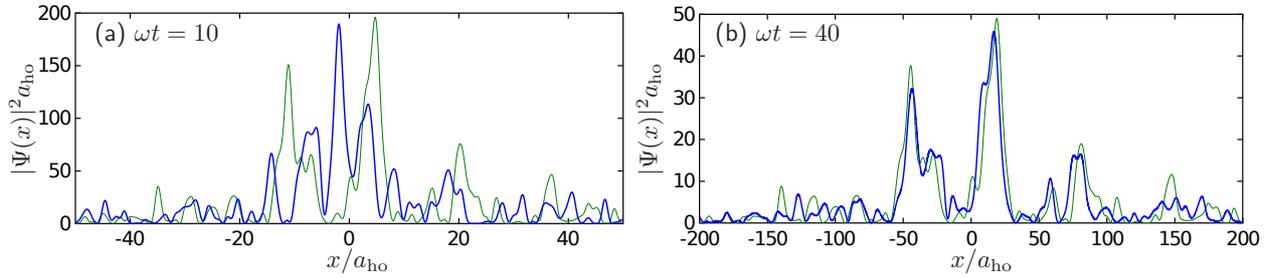}
\end{center}\vspace*{-4mm}
\caption{True density (blue) and its far-field estimate (green) given by (\ref{farfieldk}) after long times of flight. 
\label{FIG-2}}
\end{figure}

In Fig.~\ref{FIG-2}, the central section of the expanding cloud is shown for two long times $t=10/\omega$ and $t=40/\omega$. The figure also compares to the far-field estimate (\ref{farfieldk}) given by a magnified momentum distribution. The far field estimate is found wanting even at the otherwise very long time $t=10/\omega$, and becomes only passable at $t=40/\omega$. For comparison, note that the  detection time in the reference experiment \cite{Gring12} was $t_{\rm flight}=0.65/\omega\ll t_v$, which places it even well before any significant expansion.

\section{A closer look at the numerical difficulty}
\label{NUM}

Consider now calculations, e.g. in 3D, for which a lattice that properly describes both the small starting and large expanded state is extremely large.

\subsection{Computational effort}
\label{EFFORT}

Let the energy per particle in the trapped state be $\varepsilon d$, i.e. it will be $\varepsilon$ per degree of freedom in free space.
This is all converted to kinetic energy by the end of the conversion phase at $t_s$ so that a typical wavevector is $k_{\rm typical} = \sqrt{2m\varepsilon}/\hbar$. 
For good measure, and particularly to allow for energy fluctuations above the mean, one needs to include higher values $k_{\rm max} \approx r_Kk_{\rm typical}$ with $r_K\sim2$.
The spacing on the lattice needed to resolve the resulting wavelengths (Nyquist-Shannon theorem) is going to be $\Delta x_{\rm min} \le \pi/k_{\rm max} = \pi\hbar/(r_K\sqrt{2m\varepsilon})$. 
Now, when the widths of the starting cloud in the $j$th direction are $W_j$, the widths after free flight expansion will be approximately
\be\label{ulW}
\ul{W}_j=W_j + 2r_Kt_{\rm flight}\sqrt{2\varepsilon/m},
\ee
allowing again for wavevectors of up to $k_{\rm max}$. We use the underlined notation for final quantities in anticipation of a large lattice. 
The minimum number of lattice points needed in each direction for the expanded cloud is $\ul{M}_j^{\rm min} = \ul{W}_j/\Delta x_{\rm min} = (r_K/\pi\hbar)[W_j\sqrt{2m\varepsilon} + 4r_Kt_{\rm flight}\varepsilon]$.
To have an accurate calculation extra padding and resolution usually has to be included, giving $\ul{M}_j\approx r_A\ul{M}_j^{\rm min}$ with another factor $r_A\sim2$. 

After significant expansion, when the $W_j$ have become negligible, the required lattice size approaches $\ul{M}_j\approx 4r_K^2r_At_{\rm flight}\varepsilon/\pi\hbar$. 
Thus the overall required size $\ul{M} = \prod_j \ul{M}_j$ will be 
\be\label{Mnaive}
\ul{M} \ge C \left(\frac{\varepsilon\, t_{\rm flight}}{\hbar}\right)^d,
\ee
where $C= (4r_K^2r_A/\pi)^d\sim\mc{O}(10^d)$, which is about a thousand in 3D. 

Memory requirements for double precision arithmetic will be $16\times\ul{M}$ in bytes, while the time to carry out a FFT will scale as $\ul{M}\log\ul{M}$, and the time to do the evolution (\ref{free0}) is $\sim\ul{M}$. 
Defect experiments tend to have $\varepsilon t_{\rm flight}/\hbar\sim\mc{O}(100)$. For example, in the \cite{Gring12} experiment considered as an example here, $\varepsilon t_{\rm flight}/\hbar\approx 60$. 
 While the time for carrying out such an FFT on desktop computers is manageable (of the order of an hour for $\ul{M}=5\times10^9$ on one processor core), the real problem are memory requirements. For $\ul{M}=5\times10^9$, having sufficient RAM memory (75GB) on a desktop becomes troublesome.

\subsection{Information in the wavefunction}
\label{INFO}

What can be done? One can see that the effort involved in (\ref{Mnaive}) is almost all wasted because there is no new information about the cloud  in the final state $\ul{\Psi}(\ul{\bo{x}})$ that was not in the initial $\Psi(\bo{k})$. The evolution (\ref{free0}) and DFTs between x-space and k-space  (\ref{FT}), (\ref{iFT}) are deterministic and reversible. Also, we know that the final not-quite-far-field density $\ul{\rho}(\ul{\bo{x}})$ is going to be at least qualitatively similar to  (\ref{farfieldk}) which obtained via a simple magnification of the starting momentum wavefunction (see Fig.~\ref{FIG-2}). This suggests that visible structures will be much broader than in the initial state.
The trouble of course is that while the density gets magnified during the free flight by factors 
\be\label{Lambdaj}
\Lambda_j = \frac{\ul{W}_j}{W_j} =  1 + \frac{2r_Kt_{\rm flight}\sqrt{2\varepsilon}}{W_j\sqrt{m}},
\ee 
the velocity remains encoded in a wavelength that remains constant. As long as velocity information is kept, the size of the computational lattice must grow by these same factors $\Lambda_j$ to keep resolving the largely unchanged phase oscillation. The wastefulness amounts to at least a factor of $\Lambda=\prod_j\Lambda_j$. 

Clearly the thing one must do is avoid storing the entire fine lattice of size $\ul{M}$, and abandon knowledge of the properly sampled phase profile at $t_{\rm final}$, leaving only density information on a coarser lattice. 
The initial converted state $\Psi(\bo{x},t_s)$ can be fully defined on a smaller lattice with 
\be
M = \prod_j M_j = \frac{\ul{M}}{\Lambda} \approx \left(\frac{r_Ar_K\sqrt{2m\varepsilon}}{\pi\hbar}\right)^d \prod_j W_j
\ee
points and the usual spacings $\Delta\bo{x}$, which comes from (\ref{Mnaive}) and (\ref{Lambdaj}). The right hand expression assumes $\Lambda_j\gg1$. A magnification of the initial $M$ lattice by a factor of  $\Lambda_j$ in each direction, while keeping the number of points constant, should be possible in principle without adversely affecting the quality of the final density profile. 

Let us first consider an overly simple approach that tries to do this but fails in an instructive way:

\subsection{Naive DFT}
\label{NAIVE}
Since the positions appear explicitly in (\ref{iFT}), it is tempting to proceed as follows:
\begin{enumerate}
\item Obtain with the k-space wavefunction at $t_s$ represented as $\wt{\Psi}_{\wt{\bo{m}}}$ on the small $M$ lattice.
\item Apply evolution (\ref{free0}) to obtain $\wt{\Psi}(t_{\rm final})_{\wt{\bo{m}}}$ after whatever time $t_{\rm flight}$ is necessary.
\item Carry out the sum in the return transformation  (\ref{iFT}) using magnified lattice values of $\wb{\bo{x}}$ and automatically keeping the same relatively small number of points, $\wb{M}=M$. 
\end{enumerate}

An appropriate magnified lattice 
would have the same number of points as the starting state: $\wb{M}_j = M_j$, but larger spacing $\Delta\wb{x}_j = \Lambda_j\Delta x_j$, as well as  appropriately shifted zero points $\wb{a}_j$. The new positions would be 
\be\label{wbxnaive}
\wb{\bo{x}} = \wb{\bo{a}}+\bm{\Lambda}\cdot(\bo{x}-\bo{a})\qquad; \qquad \wb{x}_j = \wb{a}_j+\wb{n}_j\Delta\wb{x}_j
\ee
indexed by $\wb{n}_j=0,\dots,(M_j-1)$. 
For step 3, the discrete exponent in (\ref{iFT}) is 
\be\label{exponent}
i\bo{k}\cdot\wb{\bo{x}} = i\bo{k}\cdot\wb{\bo{a}} + i\sum_{j=1}^d\frac{2\pi}{M_j}\wt{m}_j\Lambda_j \wb{n}_j.
\ee
 To use the convenient DFT form (\ref{iDFT}), the factors $\Lambda_j \wb{n}_j$ need to be integers. Hence, the scale factor $\Lambda_j$ needs to be an integer $\lambda_j$ for all points on the final $\wb{\bo{x}}$ lattice to be calculated this way. Since phases that are a multiple of $2\pi$ are equivalent, a value of $\lambda_j\wb{n}_j > M_j$ will lead to the same $\wb{\Psi}$ as one below $M_j$. This makes any value of $\wb{n}_j$ correspond to an element of a DFT that sums over $\wt{m}_j$. Let us define an auxiliary index 
\be\label{nii}
n^{\prime\prime}_j = {\rm mod}\left[ \wb{n}_j\lambda_j , M_j \right]
\ee
dependent on $\wb{n}_j$, which indicates the element of the final inverse DFT that should to be used to obtain a particular point on the final $\wb{x}$ lattice. One obtains the following:
\be\label{Psinaive}
\wb{\Psi}^{(\rm naive)}_{\bo{n}}(t_{\rm final}) = \frac{\Lambda (2\pi)^{d/2}}{V} {\rm DFT}^{-1}\left[\wt{\Psi}_{\wt{\bo{m}}}(t_{\rm final})e^{i\wb{\bo{a}}\cdot\bo{k}_{\wt{\bo{m}}}} \right]_{\bo{n}^{\prime\prime}}.
\ee
which is very similar in form to (\ref{Psin}), except for the indexing by $\bo{n}^{\prime\prime}$, shift $\wb{\bo{a}}$ and $\Lambda$ prefactor. The last is put in by hand, to keep the amplitude of the original cloud unchanged at $t_s$ upon magnification of the lattice.
The numerical effort required by (\ref{Psinaive}) is small, with the largest matrix to be stored only of size $M$ , i.e. $\prod_j\Lambda_j$ times less than the brute force case of (\ref{Mnaive}).

\begin{figure}[htb]
\begin{center}
\includegraphics[width=\columnwidth]{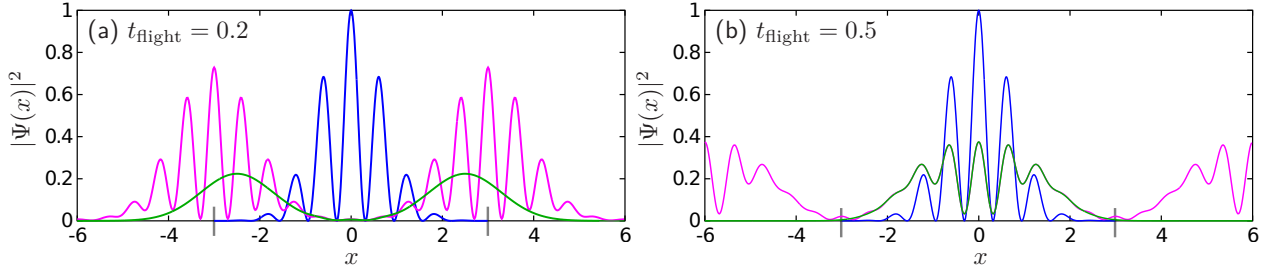}
\end{center}\vspace*{-4mm}
\caption{Initial (blue) and final (magenta)  densities after the naive attempt at expansion in 1D with prescription (\ref{Psinaive}). The initial state is $\Psi(x) = exp(-x^2/2)\cos k_0 x$ with $k_0=5$, defined on the range $x\in[-3,3]$ indicated with gray bars, with $M=600$. Expansion times are given on the plots. The true final state obtained using (\ref{Psin}) on a larger lattice defined on $x\in[-6,6]$ is shown in green. 
\label{FIG-3}}
\end{figure}

The results for a 1D test case can be seen in Fig.~\ref{FIG-3}. They are not good. This is of course because DFTs correspond to the correct Fourier transform only for the specific relationship between x- and k-space lattices that is described in Sec.~\ref{FREEFLIGHT}, and not the wishful one that step 3 implies. What is actually being carried out by (\ref{Psinaive}) instead, is the free evolution for a time $t_{\rm flight}$ of an infinite number of copies of the state $\Psi(\bo{x},t_s)$, repeated in a tiling pattern because of the periodic boundary conditions assumed by the DFT. As soon as the cloud begins to expand, it overlaps around its own edges and everything gets scrambled.

In k-space, the picture is that the flight time is so long that the phase winding caused by (\ref{free0}) advances so much that aliasing occurs. The phase difference between neighboring high-k points is $2\pi\hbar t_{\rm flight} k_{\rm max}/mL_j$ which is 
\be\label{deltatheta}
\Delta\theta_{\rm max}=\pi\left(\frac{\ul{W}_j-W_j}{L_j}\right)
\ee
after substituting for $k_{\rm max}$. 
As a result, the phase variation with $\bo{k}$ is not resolved and the state is scrambled by (\ref{Psinaive}) whenever the starting box $L_j$ is appreciably smaller than the expanded width $\ul{W}_j$. 

The moral from this naive approach is that the data in the starting  $\Psi_{\bo{n}}$ lacks the crucial physical information that the area beyond the edges of the starting lattice is supposed to be vacuum.

\section{Solving the memory problem}
\label{SOL}

\subsection{Derivation}
\label{DERIV}

To utilize the physical information about the system that the background over which the cloud expands is vacuum, let us define a buffered starting field $\Phi(\bo{r},t_s)$: 
\be\label{vacuum}
\Phi(\bo{r},t_s) =  \left\{\begin{array}{c@{\quad}l}
\Psi(\bo{r},t_s) & \text{if}\ \forall j\ :\ a_j \le r_j < (a_j+L_j)\\
0 & \text{otherwise.}
\end{array}\right.
\ee
Also, to take advantage of the highly optimized FFT algorithms, the exponent in (\ref{iDFT}) should contain only integer multiples of $2\pi i/M_j$ in each direction $j$. 
We will henceforth assume the magnification factors $\Lambda_j$ to all be positive integers $\lambda_j$, as in Sec.~\ref{NAIVE}, which suffices to obtain this condition. 
The actual cloud need not expand by an integer value, only the numerical lattice.
It may also be possible, in principle, to translate cases of fractional $\Lambda_j$ into an algorithm containing FFTs, when the $M_j$ and $\Lambda_j$ are appropriately matched. However, this would lead to combinatorial complications in the algorithm. We will refrain from considering this as it does not appear to offer any significant computational advantage. 

As usual, the general procedure to obtain a final state is to implement the time evolution in k-space as in (\ref{free0}), then use a DFT to obtain the final x-space wavefunction.  This final state is to be on the magnified lattice whose points are
\be\label{wbx}
\wb{x}_j = \wb{a}_j + \wb{n}_j\lambda_j\Delta x_j,
\ee
 lying $\Delta\wb{x}_j=\lambda_j\Delta x_j$ apart. 
$\wb{n}_j=0,\dots (M_j-1)$ is the index for the final ``small coarse'' lattice, like in (\ref{wbxnaive}).  Its volume is $\wb{V}=\prod_j\wb{L}_j$, a factor of $\lambda=\prod_j\lambda_j$ greater than the initial volume $V$. 

This last immediately presents a problem, because to obtain a final state on an x-space lattice of length $\wb{L}_j$ with an FFT requires transforming a k-space wavefunction that has resolution $2\pi/\wb{L}_j = \Delta k_j/\lambda_j$. This is $\lambda_j$ times finer than what is available in the starting state $\Psi(\bo{x})$. 
Fortunately, the vacuum assumption (\ref{vacuum}) provides sufficient information to reconstruct the fine scale structure in k-space, as we will see below.

Time evolution must also occur on this fine lattice, and to remain exact, it must not cut off high momenta, so the huge lattice $\ul{M}$ will be required, at least formally.
The required k-space wavefunction of the initial state is, generally, obtained with the transform (\ref{FT}). 
Discretizing it onto the $\ul{M}$ lattice gives 
\be\label{ex1}
\ul{\wt{\Psi}}(\ul{\bo{k}},t_s) = \frac{\Delta V}{(2\pi)^{d/2}} \sum_{\ul{\bo{x}}} \Phi(\ul{\bo{x}},t_s)\ 
e^{-i\ul{\bo{k}}\cdot\ul{\bo{x}}}.
\ee
The initial points in x-space are the $\ul{\bo{x}}$, while the k-space lattice has fine spacing $\Delta\ul{k}_j = \Delta k_j/\lambda_j$ and values $\ul{k}_j = \ul{\wt{l}}_j \Delta\ul{k}_j$ indexed by $\ul{\wt{m}}_j=0,\dots,(\lambda_jM_j-1)$ with $\ul{\wt{l}}_j={\rm mod}\left[\ul{\wt{m}}_j+\frac{1}{2}\ul{M}_j\,,\,\ul{M}_j\right]-\frac{1}{2}\ul{M}_j$. 
One in every $\lambda_j$ values of $\ul{k}_j$ will fall on a $k_j$ value that is also present in the small lattice of the starting state. In particular, instead of using the large index $\ul{\wt{\bo{m}}}$, its values can be alternatively enumerated by a pair of integers in the following way:
\be
\ul{\wt{m}}_j = \lambda_j\wt{m}_j + q_j,
\ee
where the coarse index $\wt{m}_j=0,\dots,(M_j-1)$ runs over the same set of momenta as in the starting state on the small lattice $M$, while a fine structure index $q_j = 0,\dots,(\lambda_j-1)$ counts the small $\Delta\ul{k}_j$ steps within the larger momentum step $\Delta k_j$.
The k values themselves are
\be\label{ulk}
\ul{k}_j = \Delta k_j\left(\,\wt{l}_j + \frac{q_j}{\lambda_j}\right)
\ee
when the small lattice size $M_j$ is even (as is usual). Odd $M_j$ introduces a minor but distracting complication, so from here until (\ref{iDFTbar}) we will assume even $M_j$ and return to this matter at the end of the section.
It is convenient to define a vector of fractional momentum steps  
\be
\alpha_j=\frac{q_j\Delta k_j }{\lambda_j} \in [0,\Delta k_j )
\ee
so that the fine-grained momenta can be written in a conscise vector notation:
\be\label{ulkj}
\ul{k}_j = k_j  + \alpha_j(q_j)\quad;\qquad \ul{\bo{k}} = \bo{k}_{\wt{\bo{m}}}+\bm{\alpha}_{\bo{q}}
\ee
in terms of the coarse starting momenta $\bo{k}$ and the fine grained shift $\bm{\alpha}$. 

Now luckily, the majority of the elements in the sum over points $\ul{\bo{x}}$ in (\ref{ex1}) can be discarded because they are in vacuum and contribute zero. 
Provided we make the commonsense assumption that the large lattice includes the entire lattice $\bo{x}$ for the small starting cloud, then this leaves just the sum over the usual starting state indices $\bo{n}$ defined in (\ref{xj}). With this, and substituting (\ref{xj}), (\ref{vacuum}) and (\ref{ulk}) into (\ref{ex1}), one obtains:
\be\label{ex2}
\ul{\wt{\Psi}}(\ul{\bo{k}},t_s) =
 \frac{\Delta V}{(2\pi)^{d/2}} \sum_{\bo{n}} \Psi_{\bo{n}}(t_s) 
\exp\left[-i\sum_{j=1}^d \frac{2\pi n_j}{M_j}\left(\wt{m}_j+\frac{q_j}{\lambda_j}\right)\right] e^{-i\ul{\bo{k}}\cdot\bo{a}}
\ee
which is just a sum over the small lattice. In fact, any element of the k-space wavefunction is given by an appropriate DFT on the small lattice:
\be\label{ex3}
\ul{\wt{\Psi}}(\ul{\bo{k}},t_s) = \frac{\Delta V}{(2\pi)^{d/2}}\, e^{-i\ul{\bo{k}}\cdot\bo{a}}\ 	 {\rm DFT}\left[ \Psi_{\bo{n}}(t_s)\,e^{-i\bm{\alpha}_{\bo{q}}\cdot(\bo{x}-\bo{a})} \right]_{\wt{\bo{m}}}
\ee
with the help of the coarse $\wt{\bo{m}}$ and fine $\bo{q}$ indices. 
To get the entire wavefunction, a separate FFT is required for each differing value of $\bo{q}$.

The time evolution between the DFTs is just
\be\label{time}
\ul{\wt{\Psi}}(\ul{\bo{k}},t_{\rm final}) = \ul{\wt{\Psi}}_{\wt{\bo{m}},\bo{q}}(t_{\rm final}) = \ul{\wt{\Psi}}(\ul{\bo{k}},t_s)\ \exp\left[-i \frac{\hbar t_{\rm flight} |\ul{\bo{k}}|^2}{2m}\right].
\ee

To obtain the final state in x-space, one discretizes (\ref{iFT}) and obtains the following expression on the fine lattice:
\be\label{ex4}
\ul{\Psi}(\ul{\bo{x}},t_{\rm final}) = \frac{(2\pi)^{d/2}}{\lambda V} \sum_{\wt{\bo{m}},\bo{q}} e^{i\ul{\bo{k}}\cdot\ul{\bo{x}}}\ \ul{\wt{\Psi}}_{\,\wt{\bo{m}},\bo{q}}(t_{\rm final}).
\ee
We don't need the entire huge lattice $\ul{\bo{x}}$, only the coarsened version with 
 selected sparse points given by (\ref{wbx}). 
Taking only the subset $\wb{\bo{x}}$ of points from the $\ul{\bo{x}}$ lattice  and substituting according to (\ref{wbx}) and (\ref{ulk}) gives
\be\label{ex5}
\wb{\Psi}(\wb{\bo{x}},t_{\rm final}) = \frac{(2\pi)^{d/2}}{\lambda V} \sum_{\wt{\bo{m}},\bo{q}} \ul{\wt{\Psi}}_{\,\wt{\bo{m}},\bo{q}}(t_{\rm final})\  
\exp\left[i\sum_{j=1}^d\frac{2\pi\wt{m}_j \wb{n}_j\lambda_j}{M_j}  + i\bm{\alpha}_{\bo{q}}\cdot\wb{\bo{x}} +i\bo{k}_{\wt{\bo{m}}}\cdot\wb{\bo{a}}\right].
\ee
This can almost be written as a sum of DFTs, except for one detail that was seen already in Sec.~\ref{NAIVE}: For a DFT over the small k-space lattice $\wt{\bo{m}}$ of size $M$, normally the ``x-space'' indices should run over the range $0,\dots (M_j-1)$. Here instead, in the relevant part of the exponent $i\sum_{j=1}^d\frac{2\pi\wt{m}_j \wb{n}_j\lambda_j}{M_j}$, we have the quantity $n^{\prime}_j=(\wb{n}_j\lambda_j)$ which increments in jumps: $n^{\prime}_j=0,\lambda_j,2\lambda_j,\dots,[\lambda_j(M_j-1)]$. Fortunately, the whole exponent is unchanged upon adding $M_j$ to $n^{\prime}_j$. Hence we can define the auxiliary numbers $n^{\prime\prime}_j$ like in (\ref{nii}),
which will index the output of the DFT.
Then, the final result for the coarse-grained wavefunction after flight can be written as a sum of inverse DFTs on the small lattice $M$:
\be\label{iDFTbar}
\wb{\Psi}_{\bo{\wb{n}}}(t_{\rm final}) = \frac{(2\pi)^{d/2}}{\lambda\Delta V} \sum_{\bo{q}} e^{i\bm{\alpha}_{\bo{q}}\cdot\wb{\bo{x}}}\,
{\rm DFT}^{-1}\left[ \Psi_{\wt{\bo{m}},\bo{q}}(t_{\rm final}) e^{i\bo{k}_{\wt{\bo{m}}}\cdot \wb{\bo{a}}} \right]_{\bo{n}''}.
\ee
Note how the proper expression (\ref{iDFTbar}) differs from the naive (\ref{Psinaive}) by having a sum of similar DFTs indexed by $\bo{q}$, that account for the fine-scale fractional k-space shifts $\bm{\alpha}_{\bo{q}}$. 
To calculate these, only FFTs \emph{on the small lattice $M$} are required. There is never a need to store the huge $\ul{M}$ lattice.

Finally, returning to the unusual case of odd $M_j$, (\ref{ulk}) and (\ref{ulkj}) also apply to all points except for a few with $\wt{m}_j=(M_j-1)/2$ that end up with $\ul{k}_j>k^{\rm max}_j=\pi M_j/L_j$. For these, one should substitute $\wt{l}_j\to\wt{l}'_j=(\wt{l}_j-M_j)$ in (\ref{ulk}) and $k_j\to (k_j-M_j\Delta k_j)$ in (\ref{ulkj}) to carry out the umklapp flipping to negative wavevectors on the fine momentum grid. It turns out that the only change in the intervening expressions above is that one should replace $k_j\to (k_j-M_j\Delta k_j)$ in the $\bo{k}_{\wt{\bo{m}}}$ of (\ref{ex5}) and (\ref{iDFTbar}) for the several points when $\wt{m}_j=(M_j-1)/2$ and $q_j\ge\lambda_j/2$. This actually makes very little difference in practice provided that all $\lambda_j\ll M_j$.

\subsection{The prescription}
\label{RESULT}
This is a summary of the main result that gathers the above results together. 
One starts from  the initial wavefunction $\Psi(\wb{\bo{x}},t_s)\equiv\Psi_{\bo{n}}(t_s)$ described in $d$ dimensions on an x-space lattice with $M_j$ points in each dimension  $j=1,\dots,d$ of length $L_j$. The region outside this lattice is initially vacuum. The positions of the points are
\be
x_j = a_j + \frac{n_j L_j}{M_j}
\ee
with indices $n_j=0,\dots,(M_j-1)$. 
The momentum spacing is $\Delta k_j=2\pi/L_j$. 
Free flight occurs for a time interval $t_{\rm flight}=t_{\rm final}-t_s$. Subsequently the lattice spacing on which the wavefunction is described in x-space is magnified by integer factors $\lambda_j$, giving lattice points
\be
\wb{x}_j = \wb{a}_j + \frac{\wb{n}_j \lambda_jL_j}{M_j}
\ee
with indices $\wb{n}_j=0,\dots,(M_j-1)$. 
The volume of the initial lattice is $V=\prod_jL_j$, the number of points $M=\prod_jM_j$, the volume magnification $\lambda=\prod_j\lambda_j$. 
The fractional momentum steps
\be
\alpha_j=\frac{\Delta k_j q_j}{\lambda_j} \in [0,1)\Delta k_j
\ee
form a vector $\bm{\alpha}_{\bo{q}}$ that is enumerated by the indices $q_j=0,1,\dots,(\lambda_j-1)$. Bold quantities are vectors in $d$-dimensional space.
The final x-space wavefunction on the magnified lattice is given by:
\begin{subequations}
\label{FF}
\bea
\label{FFpsi}
\wb{\Psi}_{\wb{\bo{n}}}\,(t_{\rm final}) &=& \sum_{\bo{q}} f^{(\bo{q})}_{\wb{\bo{n}}}\ B^{(\bo{q})}_{\bo{n}''}\\
\label{presf}
f^{(\bo{q})}_{\wb{\bo{n}}} &=& \frac{1}{\lambda} \exp\left[i\bm{\alpha}_{\bo{q}}\cdot\left(\wb{\bo{x}}_{\bo{\wb{n}}}-\bo{a}-\bm{\alpha}_{\bo{q}}\frac{\hbar t_{\rm flight}}{2m}\right)\right]\\
\label{presB}
B^{(\bo{q})}_{\bo{p}} &=& {\rm FFT}^{-1}\left[ \wt{A}^{(\bo{q})}_{\wt{\bo{m}}} e^{i\,\bo{k}_{\wt{\bo{m}}}\cdot\,\left[ \wb{\bo{a}}-\bo{a} - \left(\bo{k}_{\wt{\bo{m}}} +2\bm{\alpha}_{\bo{q}}\right)\frac{\hbar t_{\rm flight}}{2m}\right]}
\right]_\bo{p}\\
\label{presA}
\wt{A}^{(\bo{q})}_{\wt{\bo{m}}} &=& {\rm FFT}\ \left[ \Psi_{\bo{n}}(t_s) e^{-i\bm{\alpha}_{\bo{q}}\cdot\left[\bo{x}-\bo{a}\right]}
\right]_{\wt{\bo{m}}}
\eea
when all $M_j$ are even, with the elements of the  auxilliary index in (\ref{FFpsi}) being
\be\label{FFaux}
n^{\prime\prime}_j = {\rm mod}\left[\wb{n}_j\lambda_j,M_j\right],
\ee
and FFT indicating fast Fourier transforms. The wavevectors $\bo{k}_{\wt{\bo{m}}}$ are given by (\ref{kj}).
\end{subequations}

If any $M_j$ are odd, then in the $\bo{k}_{\wt{\bo{m}}}$ of (\ref{presB}) one should further umklapp the maximum k value in each of those dimensions $j$ when $q_j\ge\lambda_j/2$. That is, for $\wt{m}_j=(M_j-1)/2$ and $q_j\ge\lambda_j/2$, replace $k_j(\wt{m}_j)$ by $-\Delta k_j (M_j+1)/2$.

\section{3D Example}
\label{3DEX}

\begin{figure}[htb]
\begin{center}
\includegraphics[width=\columnwidth]{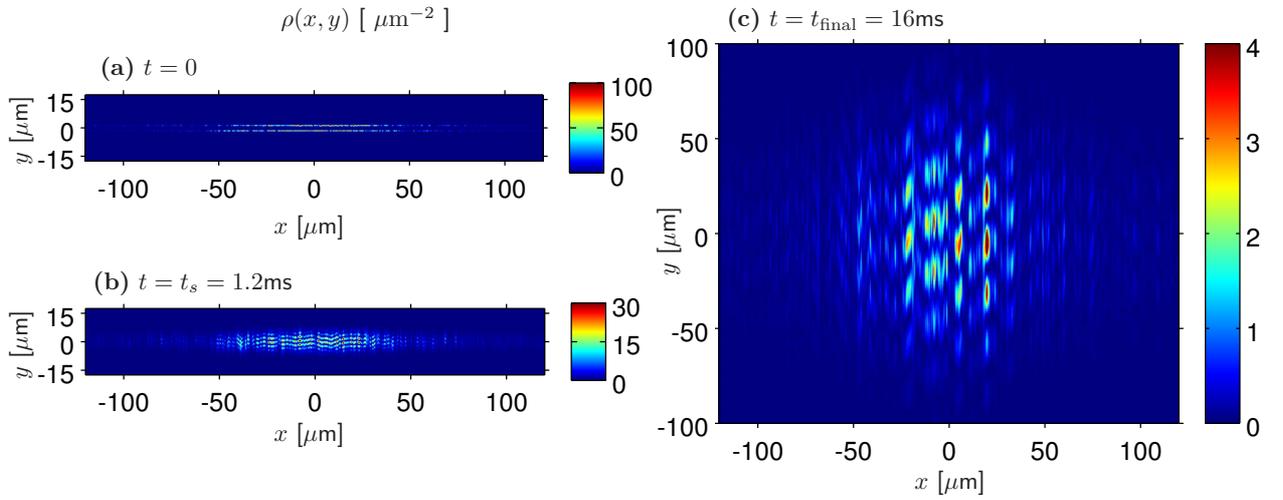}
\end{center}\vspace*{-4mm}
\caption{
Density patterns during the free flight described in Sec.~\ref{3DEX}. Shown is the $x,y$ density $\rho(x,y) = \int dz |\Psi(\bo{r})|^2$. 
Panel (a): initial state released from the traps at $t=0$. 
Panel (b): The ``starting'' state for the free expansion, at  $t=t_s=1.2$ms, after the conversion phase.
Panel (c): The state at the detector at $t=t_{\rm final}=16$ms.
\label{FIG-4}}
\end{figure}

As an example of a calculation that cannot be done by brute force on a PC, consider a full simulation of the relative phase measurements in the Vienna experiment \cite{Gring12}.
Here, two neighbouring ultracold atomic clouds of ${}^{87}$Rb, $\Psi^{(\pm)}$, are initially populated in quasi-one-dimensional harmonic traps that are elongated in the $x$ direction, as seen in Fig.~\ref{FIG-4}(a). Trap frequencies are $\omega=2\pi\times 6.4$ Hz in the $x$ direction, and $\omega_{\perp}=2\pi\times1400$ Hz in the transverse directions. The clouds are initially separated by a small gap of $D=2.75$nm$ = 0.645a_{\rm ho}$ in the $y$ direction. The traps are released  at $t=0$, and rapid expansion takes place in the $y$ and $z$ directions. The clouds soon interfere, forming a fringe pattern, as shown in (Figs.~\ref{FIG-4}(b-c))). The local displacement $\delta y(x)$ of the fringes in the $y$ direction from the middle position $y=0$ corresponds to the phase difference that existed locally between the two initial clouds at $t=0$. 
\be\label{dphase}
\delta y(x) \propto \Delta\theta_0(x) = \angle\Psi_0^{(+)}(x) - \angle\Psi_0^{(-)}(x) 
\ee 
The fringe pattern is detected after 16ms of free flight ($t_{\rm final}=0.65/\omega$) at a detector. This is basically an integral of the final density over the $z$ direction.
In this way, a local phase difference measurement on the initial clouds can be made.

There is of course some distortion during flight, as was seen in Sec.~\ref{1DEX}, and a simulation of the expansion can be necessary to see quantitatively how the pattern at the detector corresponds to the initial phase profile. 
Here the case where the two clouds are populated by independent thermal gases will be considered. The one-dimensional c-field wavefunctions $\Psi_{\rm ic}^{(\pm)}(x)$ are generated using the same SGPE method as in Sec.~\ref{1DEX} and the same parameters. Eq.~(\ref{SGPE}) is run separately with independent noises for each of the two clouds. 
As these are quasi-1D traps, the wavefunction in the $y$ and $z$ directions is well approximated by just the harmonic oscillator ground state. 
The initial state (see Fig.~\ref{FIG-4}(a)) consists of two terms: 
\be\label{3dic}
\Psi_0(\bo{x}) = \sqrt{\frac{m\omega_{\perp}}{\pi\hbar}}
\sum_{\pm} \Psi_{\rm ic}^{(\pm)}(x)\exp\left[-\left\{\,\left(y\pm \frac{D}{2}\right)^2+z^2\right\}\frac{m\omega_{\perp}}{2\hbar}\right].
\ee

The conversion phase is simulated with the 3D Gross-Pitaevskii equation (GPE) 
\be\label{GPE}
\hbar\frac{d\,\Psi(\bo{x},t)}{\partial t} = -i\left[-\frac{\hbar^2}{2m}\nabla^2 + g\,|\Psi(\bo{x},t)|^2\right]\Psi(\bo{x},t)
\ee
using with a semi-implicit split-step algorithm \cite{Drummond90}. It is run
until 1.2ms, which is $t_s=0.05/\omega$.
The s-wave scattering length is 5.24nm, giving a value of $g=0.01544\hbar\omega a_{\rm ho}^3$ for the interaction strength in 3D in terms of $a_{\rm ho}=\sqrt{\hbar/m\omega}$. 
The lattice used has dimensions $L_x = 67.5a_{\rm ho}, L_{y,z} = 8.26a_{\rm ho}$, and $M=2304\times256\times256$ points. The calculation took 1hour 20 mins on an Intel 2.4 GHz CPU using the FFTW library \cite{FFTW05}, and used 8\% of the 96GB RAM memory on the PC.
The situation at $t_s$ is shown in Fig.~\ref{FIG-4}(b). 

Subsequently, the $\Psi(\bo{x},t_s)$ were fed into the free flight prescription (\ref{FF}) developed here, for expansion out to the detector at 16ms. The lattice expansion factors were $\lambda_y=\lambda_z=8$ and $\lambda_x=1$, i.e. no expansion in the $x$ direction. However, the initial $\wb{x}$ lattice was slightly buffered with vacuum at the edges with respect to the one used for the conversion phase, having 
a length $\wb{L}_x=94.92a_{\rm ho}$. This was to allow some natural spreading, which was too small to make a lattice expansion by $\lambda_x=2$ worthwhile. The final coarse lattice had $\wb{M} = 3240\times256\times256$ points. The calculation took 64 minutes on the same PC and used 14\% of RAM. 
The resulting predicted detector image is shown in Fig.~\ref{FIG-4}(c).

For comparison, a brute force calculation using the plain (\ref{Psin}) was not able to reach the detection time. The best that was obtained on the aforementioned PC without going into swap space was expansion out to $t=0.40/\omega$, corresponding to 10ms, or 62\% of the flight. This took 3 hours on a $\ul{M}=2880 \times 1350 \times 1350$ lattice, and used 88\% of RAM, as well as requiring special additional work with the code to pass 64 bit pointers into the FFTW library. 64 bit pointers were required when $\ul{M}\ge2^{31}$. 

\begin{figure}[htb]
\begin{center}
\includegraphics[width=0.5\columnwidth]{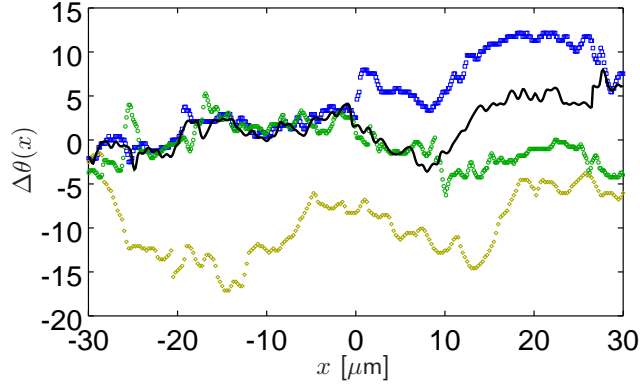}
\end{center}\vspace*{-4mm}
\caption{Phase differences along $x$ in the central region of the cloud, after unwrapping modulo $2\pi$ to remove sudden jumps greater then $\pi$. The black line shows the actual phase difference $\angle\Phi_+(x) - \angle\Phi_-(x)$  between the initial clouds. 
The blue squares and green circles show the apparent phase differences inferred from the fringe shifts $\delta y(x)$ in the density images of Fig.~\ref{FIG-4}(b,c), respectively, using (\ref{dphase2}).
The yellow diamonds show what would be inferred from the very late density image of Fig.~\ref{FIG-6} at 62ms. 
\label{FIG-5}}
\end{figure}

Fig.~\ref{FIG-5} shows the apparent phase differences that can be inferred from the free-expansion at different times, and the true initial phase difference. 
The fringe shift $\delta y(x)$ was estimated from the $y$ position of the maximum density peak for a given $x$.
The proportionality constant in (\ref{dphase}) can be estimated by considering the free flight evolution of (\ref{3dic}) in the $y$ direction only, ignoring other effects. One finds
\be
\psi(y,t) = \frac{1}{\sqrt{1+i\omega_{\perp}t}}\sum_{\pm} A_{\pm}\exp\left[-\frac{m\omega_{\perp}(2y\mp D)^2}{8\hbar(1+i\omega_{\perp}t)}\right]. 
\ee
where $A_{\pm} = \left(\frac{m\omega_{\perp}}{\pi\hbar}\right)^{1/4}\Psi_0^{(\pm)}(x)$. We are most interested in the limit $\omega_{\perp}t\gg1$ and $y\gg D$. We expand the exponents in the density $|\psi(y)|^2$ to lowest nontrivial orders
in the small quantities $\eta_t=1/\omega_{\perp}t$ and $\eta_y=D/y$: that is, $\mc{O}(\eta_t^2,\eta_y^{-2})$ for the amplitude and $\mc{O}(\eta_t,\eta_y^{-1})$ for the phases. This gives
\be
|\psi(y,t)|^2 \approx
 \frac{\exp\left[-\frac{2my^2}{\hbar\omega_{\perp}t^2}\right]}{\sqrt{1+(\omega_{\perp} t)^2}}
\left\{\,
|A_+|^2+|A_-|^2+2|A_+|A_-|\cos\left[\frac{my D}{\hbar t} - \Delta\theta_0(x)\right]\,\right\}.
\ee
Hence, the phase difference estimate (modulo $2\pi$) is 
\be\label{dphase2}
\Delta\theta_0(x) \approx y_{\rm peak}(x)\ \frac{m D}{\hbar t}.
\ee 
where $y_{\rm peak}$ is the location of the peak nearest to $y=0$. 
What we see in Fig.~\ref{FIG-5} is that the global long-wavelength behaviour of the phase difference is generally predicted well by the fringes in the expanded cloud. This is apart from some remnant localized shifts of modulo $2\pi$ that move an entire segment by $2\pi$ without affecting the long-wavelength phase trend. These shifts are at 0$\mu$m and 28$\mu$m for the early cloud at $t_s$, and near 10$\mu$m for the cloud at the detector. However, one can also see that true to the behaviour seen in the 1D case of Sec.~\ref{1DEX}, the prediction of local details in the phase difference is largely scrambled during the time of flight.

\begin{figure}[htb]
\begin{center}
\includegraphics[width=0.5\columnwidth]{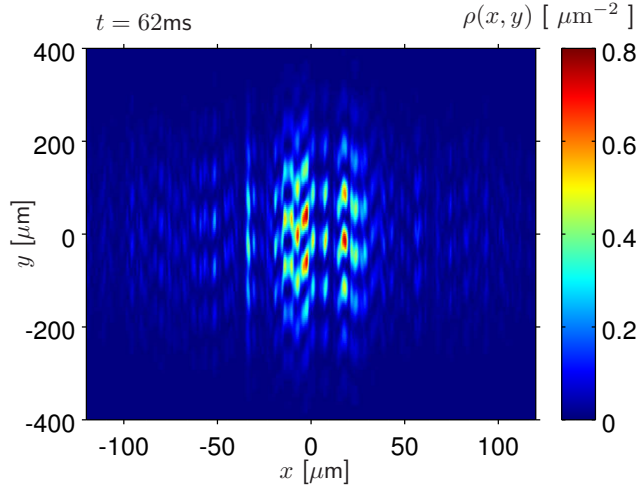}
\end{center}\vspace*{-4mm}
\caption{
The density pattern that would be seen after $t=62$ms of free flight, details as in Fig.~\ref{FIG-4}, except for a 1:4 aspect ratio between the axes.
\label{FIG-6}}
\end{figure}

The effort required by (\ref{FF}) for even very long expansions scales relatively graciously. For example, continuing the expansion out to $t=62$ms$ = 2.5/\omega$ is also possible. 
This is shown in Fig,~\ref{FIG-6} and the yellow plot in ~\ref{FIG-5}, and one sees continuing change in the fringe profile. In particular, the phase difference estimate is starting to become bad, with large long-scale discrepancies.
This because now the expansion has lasted long enough that a lot of movement of the defects in the  $x$ direction has occurred. This is expected, since the estimate for formation time of a momentum distribution from Sec.~\ref{DEFECTS} is $t_v=\sqrt{8}/\omega=70$ms here. The final cloud is now a relatively huge 
$0.3\times 0.6\times0.6$ mm 
in size, having expanded by a factor of about $10^4$ in the $y$ and $z$ directions since release from the trap. The final lattice is scaled by 
$\lambda_x=2$, and $\lambda_y=\lambda_z=30$ 
from that at $t_s$. This calculation took 
18 hours on the reference PC, and used 7GB of memory. The direct approach with the $\ul{M}$ lattice would have needed 4000GB.

\section{Discussion}
\label{DISCUSSION}

\subsection{Efficiency}
\label{EFF}

To take advantage of the memory savings in (\ref{FF}), one should evaluate each term in the sum (\ref{FFpsi}) labeled by $\bo{q}$ sequentially, and accumulate its contribution to $\wb{\Psi}_{\wb{\bo{n}}}(t_{\rm final})$. 
This way, memory requirements will be $\approx2M$ complex numbers [32$\times M$ bytes for the usual double precision] -- one array of size $M$ for carrying out the FFTs, and one to store the accumulated sum $\wb{\Psi}_{\wb{\bo{n}}}(t_{\rm final})$. Some time efficiency can be gained by using a third array of size $M$ to store the starting state $\Psi_{\bo{n}}(t_s)$, but using 48$\times M$ bytes in total.

The computational load in terms of operations scales as
\be\label{cpuload}
\lambda M\log M
\ee
which is slightly faster than the brute force approach that would use (\ref{Psin}) directly on a vacuum padded lattice $\ul{M}$ and take $\sim\lambda M\log(\lambda M)$ operations.
The speed-up is mostly marginal -- an improvement by a   factor of $(1+\log\lambda/\log M)$.
The somewhat surprising result that there is any speed up at all compared to the highly-optimized FFT on $\ul{M}$ is due to the fact that so much of the initial system is vacuum and does not contribute. 
For this reason, there is no advantage to be gained by trying to use the maximum starting lattice size  $M$ that will fit in memory (perhaps after padding the nonzero part of the field that comes out of the conversion phase with vacuum), and minimum magnification $\lambda$.

The memory needed is of course strongly reduced -- by a factor of at least $\lambda/2$, comparing to the most memory-efficient in-place FFT on the huge $\ul{M}$ lattice. 

An expansion in $d$ directions requires summing of a number of terms that grows as $(t_{\rm flight})^d$. This can eventually become fairly time intensive as was seen for the calculation of Fig.~\ref{FIG-6}. 
Memory use never budges above the baseline no matter how long the flight takes.

The time needed can be alleviated by an extremely basic parallelization. Namely, distributing the evaluation of the $B^{(\bo{q})}$ on many processing cores. Up to $\lambda$ cores could be used to obtain the result in a time $\sim M\log M$. However, a significantly smaller number will be optimal since $\lambda$ FFTs in parallel will require $\sim\lambda M$ numbers stored in memory again, which is what one is trying to avoid.

\subsection{Relationship to fast Fourier transforms}
\label{FFT}

The algorithm presented in (\ref{FF}) bears some rough resemblance to a Cooley-Tukey FFT algorithm \cite{Cooley65}  with radix-$\lambda_j$. The similarity is that the end results of the FFTs on the smaller $M$ lattice are multiplied by twiddle factors in (\ref{presf}). These  involve $e^{i\alpha_j\wb{x}_j}$ that introduces fractional phase shifts compared to those available on the FFT lattice.  However, the overall procedure is quite different to Cooley-Tukey and relies heavily on the vacuum padding assumption (\ref{vacuum}). This allows it to e.g. perform the two sequential FFTs on each $\bo{q}$-th term in the final sum (\ref{FFpsi}). This is also what allows the effort to scale as $\lambda$ instead of the $\lambda^2$ that would be expected from a manual summation of smaller $M$-size FFTs. 

Looking at (\ref{FFaux}), one can see that when  $\lambda$ and $M$ have common factors, this $n''$ index only accesses a part of the field $B^{(q)}_{\bo{p}}$. 
One can try to gain some computational advantage from this by using a pruned FFT \cite{Sorensen93} for the (\ref{presB}) step, to calculate only the required $\bo{p}$ values.  The advantage of pruned FFTs is not huge though. This would reduce the overall computational effort at most from $2M\log M$ to $2M\log M-M\log\lambda$ .

\subsection{Loss of phase information}
\label{COMP}

\begin{figure}[htb]
\begin{center}
\includegraphics[width=\columnwidth]{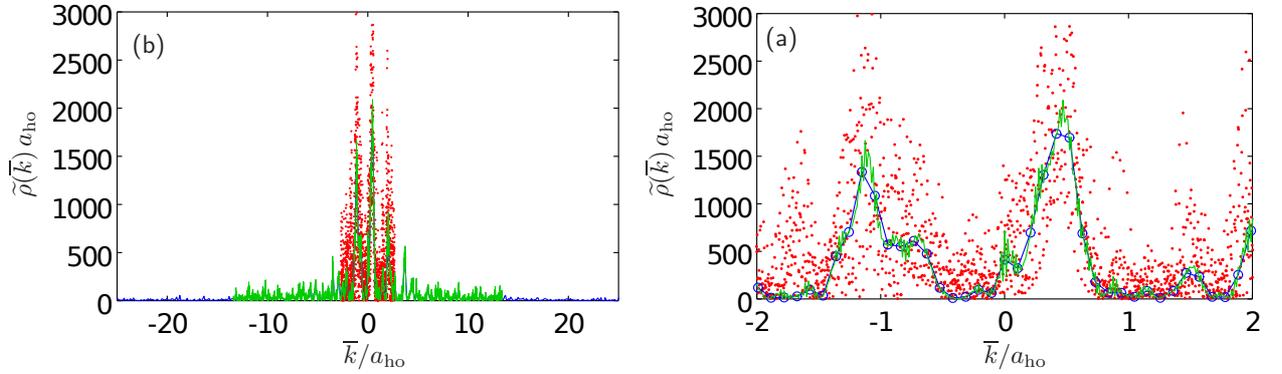}
\end{center}\vspace*{-4mm}
\caption{Loss of momentum information in the final wavefunction generated by (\ref{FF}). The plot shows: 
In blue: the true k-space density $\wt{\rho}(k)=|\wt{\Psi}(k)|^2$ in the cloud on the initial ($M$) lattice. Physically, this is preserved during free evolution;
In green and red:  The apparent k-space density $\wt{\rho}(\wb{k})=|\wt{\wb{\Psi}}(\wb{k})|^2$ , inferred by a DFT (\ref{badDFT}) of the coarse grained final wavefunction $\wb{\Psi}(\wb{x},t_{\rm final})$ in x-space. 
Green is for a magnification of $\lambda=8$ at $t_{\rm flight}=10/\omega$, while red is for a longer time of $t_{\rm flight}=40/\omega$ with $\lambda=40$. Other parameters as in Sec.~\ref{1DEX}. 
Panel (b) is a magnification of part of Panel (a). 
\label{FIG-7}}
\end{figure}

An important feature of the algorithm (\ref{FF}) to be aware of is that while the density in x-space at $t_{\rm final}$ is calculated precisely, the $\wb{\Psi}_{\wb{\bo{n}}}(t_{\rm final})$ is not generally viable for further evolution, 
and does not store the correct momentum distribution. This is because of the phase aliasing (\ref{deltatheta}) discussed in Sec.~\ref{NAIVE}. 

The wavefunction that can be reconstructed from $\wb{\Psi}_{\wb{\bo{n}}}(t_{\rm final})$ is:
\be\label{badDFT}
\wt{\wb{\Psi}}_{\wt{\wb{\bo{m}}}} = \frac{\lambda\Delta V}{(2\pi)^{d/2}}\ e^{-i\wb{\bo{a}}\cdot\wb{\bo{k}}_{\wt{\wb{\bo{m}}}}}\ {\rm DFT}\left[\wb{\Psi}_{\bo{n}}\right]_{\wt{\wb{\bo{m}}}}.
\ee
This sits on a fine k-space lattice $\wb{k}_j(\wt{\wb{m}}_j) = \wt{\wb{l}}_j\,(2\pi/\wb{L}_j)$ with $\wt{\wb{l}}_j={\rm mod}\left[\wt{\wb{m}}_j+\frac{1}{2} M_j\,,\,M_j\right]-\frac{1}{2}M_j$. 
The resulting momentum distribution is shown in Fig.~\ref{FIG-7}, for the same 1D system that was studied in Sec.~\ref{1DEX}. This time, the initial wavefunction $\Psi_{\rm ic}(x)$ was evolved to times $t_{\rm flight}$ using the prescription (\ref{FF}) on the initial $M=2048$ lattice rather than the standard step-by-step evolution  (\ref{freex})  on $\ul{M}=81920$ that was used in Sec.~\ref{1DEX}.
The green case at $t_{\rm flight}=10/\omega$  might still be passable for some purposes, though the high momenta are already lost. The red longer-time case is completely scrambled.

The fact that the phase structure in x-space remains small-scale despite a magnification of the density stymies several superficially promising ideas on how to increase the efficiency of the expansion calculation:

First, one could be tempted to try to reduce the processing load to only $\sim \log\lambda$ FFTs on $M$-points instead of the present $\lambda$ FFTs, by implementing several sequential expansions (\ref{FF}) by small factors, say $\lambda_j=2$.
However, at each such step we are left with a discretized wavefunction that has has its momentum-space tails truncated. This will soon come to resemble the red line in Fig.~\ref{FIG-7} and become useless for further evolution. 

Another approach that has been discussed in the field would try to introduce a time-dependent lattice spacing $\Delta\bo{x}(t) = F(t)$ that would track the expected density structure
while keeping the lattice size $M$ constant. The hope is that it would allow one to keep the full nonlinear equation (\ref{GPE}) at the cost of some additional correction terms dependent on $F(t)$.
 However, one can see that this will be unsuccessful as soon as the phase structure becomes too fine for the growing $\Delta\bo{x}(t)$.

\subsection{Generalizations}

The algorithm is readily adapted to cases where several complex-valued fields are present. One such case that may be aided with the algorithm presented here are positive-P simulations of supersonic BEC collisions \cite{Deuar07,Deuar11,Kheruntsyan12,Lewis-Swan15}. Here, two independent complex-valued fields $\psi(\bo{x})$ and $\psi^+(\bo{x})$ that correspond to the $\op{\Psi}(\bo{x})$ and $\dagop{\Psi}(\bo{x})$ Bose fields are used, and allow for the exact treatment of quantum fluctuations. 
The comparison of calculated and experimental pair velocity correlation widths has been problematic in these systems , because of the narrowness of the correlation peak in velocity \cite{Kheruntsyan12}. The detected peak is distorted in comparison with its k-space prediction due to not yet being in the far-field regime. There is no hope of a direct calculation of the free flight because the quantity $\varepsilon t_{\rm flight}/\hbar$ of (\ref{Mnaive}) is very high (up to $\sim10^4$) in BEC collision experiments. 

The approach can also be trivially adapted to cases of other spectra than the free particle one. This simply requires a modification of (\ref{free0}) to 
$\wt{\Psi}(\bo{k},t_{\rm final}) = \wt{\Psi}(\bo{k},t_s) \exp\left[-it_{\rm flight}\omega_\bo{k}\right]$, 
with appropriate tweaks in (\ref{presf}) and (\ref{presB}). The crucial element is the presence of the vacuum assumption (\ref{vacuum}).

\subsection{Conclusions}
\label{CONC}
To conclude, an algorithm (\ref{FF}) has been presented 
that allows the exact calculation of the density of a wavefunction freely expanding into vacuum for practically arbitrary flight times 
without filling up the computer memory. The memory requirements do not depend on flight time and are the same size as the initial input state. Computation time is slightly faster than using an FFT on a large vacuum padded lattice. 
It is implemented using standard FFT libraries and some summing of terms.
The approach relies crucially on two physical inputs: (1) That the initially compact wavefunction expands into vacuum, and (2) that the density length scale of the expanded cloud grows approximately linearly with time.
The approach makes no assumptions about symmetries of the system or about the input wavefunction, so that it is a black box tool that can be immediately applied to general cases.
This makes it well suited to the study of wavefunctions containing defects or samples of a thermal ensemble, a topic of many recent experiments \cite{Gring12,Chomaz15,Serafini15,Lamporesi13,Donadello14,Sadler06,Weiler08}. The flight times over which nontrivial defect evolution occurs during free flight are estimated in Sec.~\ref{KINETIC}.

\section*{Acknowledgments}
  I would like to thank Ray Ng and Mariusz Gajda 
for helpful discussions on this matter, and Julius de Hond whose feedback identified an omission in the manuscript.   
The work was supported by the National Science Centre (Poland) grant No. 2012/07/E/ST2/01389.

\section*{References}
\bibliography{TOF}

\end{document}